\documentclass[12pt]{article}
\usepackage{graphicx}
\usepackage{caption}
\usepackage{subcaption}
\pdfoutput=1

\newcommand\pubnumber{DPF2015-193}
\newcommand\pubdate{\today}

\def\iit{Department of Physics\\
Life Sciences\\
3101 S. Dearborn St., Room 182\\
Illinois Institute of Technology\\
Chicago, IL 60616}
\def\support{\footnote{Presenter: kgilje@iit.edu}}

\def\Title#1{\begin{center} {\Large #1 } \end{center}}
\def\Author#1{\begin{center}{ \sc #1} \end{center}}
\def\Address#1{\begin{center}{ \it #1} \end{center}}

\newcommand\pubblock{\rightline{\begin{tabular}{l} \pubnumber\\
         \pubdate  \end{tabular}}}
\newenvironment{Abstract}{\begin{quotation}  }{\end{quotation}}
\newenvironment{Presented}{\begin{quotation} \begin{center} 
             PRESENTED AT\end{center}\bigskip 
      \begin{center}\begin{large}}{\end{large}\end{center} \end{quotation}}
\def\Acknowledgments{\bigskip  \bigskip \begin{center} \begin{large}
             \bf ACKNOWLEDGMENTS \end{large}\end{center}}

\def\beq{\begin{equation}}
\def\eeq#1{\label{#1}\end{equation}}
\def\eeqn{\end{equation}}

\def\beqa{\begin{eqnarray}}
\def\eeqa#1{\label{#1}\end{eqnarray}}
\def\eeqan{\end{eqnarray}}

\let\bar=\overbar

\def\Dslash{\not{\hbox{\kern-4pt $D$}}}
\def\dslash{\not{\hbox{\kern-2pt $\del$}}}

\def\msb{{\bar{\ssstyle M \kern -1pt S}}}

\begin{document}
\begin{titlepage}
\pubblock

\vfill
\Title{Sensitivity and Discovery Potential of the PROSPECT Experiment}
\vfill
\Author{ Karin Gilje\support \\on behalf of the PROSPECT Collaboration}
\Address{\iit}
\vfill
\begin{Abstract}
Measurements of the reactor antineutrino flux and spectrum compared to model predictions have revealed an apparent deficit in the interaction rates of reactor antineutrinos and an unexpected spectral deviation. 
PROSPECT, the Precision Reactor Oscillation Spectrum measurement, is designed to make a precision measurement of the antineutrino spectrum from a research reactor and search for signs of an eV-scale sterile neutrino. 
PROSPECT will be located at the High Flux Isotope Reactor (HFIR) at Oak Ridge National Laboratory and make use of a Highly Enriched Uranium reactor for a measurement of the pure U-235 antineutrino spectrum. 
An absolute measurement of this spectrum will constrain reactor models and improve our understanding of the reactor antineutrino spectrum. 
Additionally, the planned 3-ton lithium-doped liquid scintillator detector is ideally suited to perform a search for sterile neutrinos. 
This talk will focus on the sensitivity and discovery potential of PROSPECT and the detector design to achieve these goals.
\end{Abstract}
\vfill
\begin{Presented}
DPF 2015\\
The Meeting of the American Physical Society\\
Division of Particles and Fields\\
Ann Arbor, Michigan, August 4--8, 2015\\
\end{Presented}
\vfill
\end{titlepage}
\def\thefootnote{\fnsymbol{footnote}}
\setcounter{footnote}{0}

\section{Introduction}

Describing and understanding the flavor oscillation between neutrinos remains a primary goal in the field of neutrino physics.
Many types of experiments have been conducted using neutrinos from several different sources.
One such source is nuclear reactor cores  which produce antineutrinos from the fission processes that occur within them.

The general method to detect antineutrinos coming from a reactor is through the mechanism of Inverse Beta Decay (IBD):

\begin{equation}
\bar{\nu}_e + p \rightarrow e^+ + n.
\end{equation}

Typically, a prompt signal, closely related to the antineutrino energy, is detected from the positron slowing and annihilation and a delay signal is detected through the capture of the free neutron on a nucleus such as lithium, gadolinium or hydrogen.

In order to use the antineutrino emissions of nuclear reactors for oscillation studies or other various applications, accurate predictions of the rate and the spectral shape are required.
There are two common methods used to model reactor antineutrino emissions: \textit{ab initio} and $\beta$-spectrum conversion.
The \textit{ab initio} approach calculates the expected neutrino spectrum by using fission daughter product yields and $\beta$-branch information obtained from nuclear databases such as the Evaluated Nuclear Structure Data File (ENSDF).
The $\beta$-spectrum conversion method examines the $\beta$ spectrum measured in dedicated experiments and uses unfolding techniques to infer the associated $\bar{\nu}_e$ spectrum \cite{Schreckenbach:1985}.
For commercial cores, a predicted reactor antineutrino spectrum must be known separately for all four primary fission isotopes (${}^{235}\textrm{U}$, ${}^{238}\textrm{U}$, ${}^{239}\textrm{Pu}$, and ${}^{241}\textrm{Pu}$).
For the most recent suite of $\theta_{13}$ experiments, the pure conversion approach of Huber and a hybrid method of Mueller have received the most attention \cite{Huber:2011, Mueller:2011}.
A recently produced \textit{ab initio} prediction from Dwyer and Langford has also received significant attention \cite{Dwyer:2014}.

Two issues have emerged: there is a 10\% excess in the number of observed events around 5 MeV in prompt energy and there is an overall 5\% deficit in the total number of observed neutrino events, relative to model predictions in recent experiments.
The modeling approaches are necessarily approximate, and rely upon incomplete nuclear data.
Thus, one solution to the flux and spectrum anomalies is that important nuclear processes or $\beta$-decay data are being inappropriately neglected in the reactor emission models.
Given the great complexity of the fission reactor system, direct, high-precision measurements of the reactor antineutrino spectrum provide an efficient path to greatly improving our understanding of these emissions \cite{Hayes:2015}.
An alternate explanation for the apparent deficit in the rate of observed neutrinos is the existence of an eV-scale sterile neutrino, $\nu_s$, that does not interact via the weak interaction.
However, active neutrinos can still oscillate to a sterile state leading to the overall effect of a decrease in the rate of observed events.
There are several different anomalies that could be understood with the introduction of a new neutrino beyond the Standard Model at a $\Delta m^2 \approx 1 \textrm{eV}^2$ such as the LSND, MiniBooNE and Gallium experiments \cite{Kopp:2013}.
Short baseline reactor experiments specifically measure the neutrino mixing matrix element $|U_{e4}|^2$ through $\bar{\nu}_e$ disappearance measurements which are complementary to $\nu_\mu$ disappearance measurements like those planned for the Fermilab Short Baseline Neutrino (FSBN) program which measures a mixture of $|U_{e4}|^2$ and $|U_{\mu4}|^2$.

\section{The PROSPECT Design}

PROSPECT, the Precision Reactor Oscillation and Spectrum experiment, has two goals: make a precise measurement of the ${}^{235}\textrm{U}$ reactor antineutrino spectrum and perform a sterile neutrino search focusing on the parameter space around $\Delta m^2 \approx 1 \textrm{eV}^2$.

PROSPECT will be located near the High Flux Isotope Reactor (HFIR) at Oak Ridge National Laboratory (ORNL).
HFIR is a compact-core research reactor that operates at 85 MW.
The core is a cylinder with a radius of 0.25 m and a height of 0.5 m.
It is one of the last remaining Highly Enriched Uranium (HEU) fuel research reactors in the US in which nearly all fissions come from ${}^{235}\textrm{U}$.
Most power or commercial reactors are a mixture of the four main fuel types listed above and are termed Low Enriched Uranium (LEU) reactors.
This will enable PROSPECT to cleanly extract the ${}^{235}\textrm{U}$ spectrum and help future LEU experiments produce better constrained predictions for other future oscillation measurements \cite{Juno:2015}.
HFIR runs for 41\% of the year allowing for in-depth studies of cosmogenic backgrounds during reactor-off periods.

\begin{figure}[htb]
\centering
\begin{subfigure}[b]{0.35\textwidth}
\includegraphics[clip=true, trim=2.5in 1.5in 2.5in 1in, width=\textwidth]{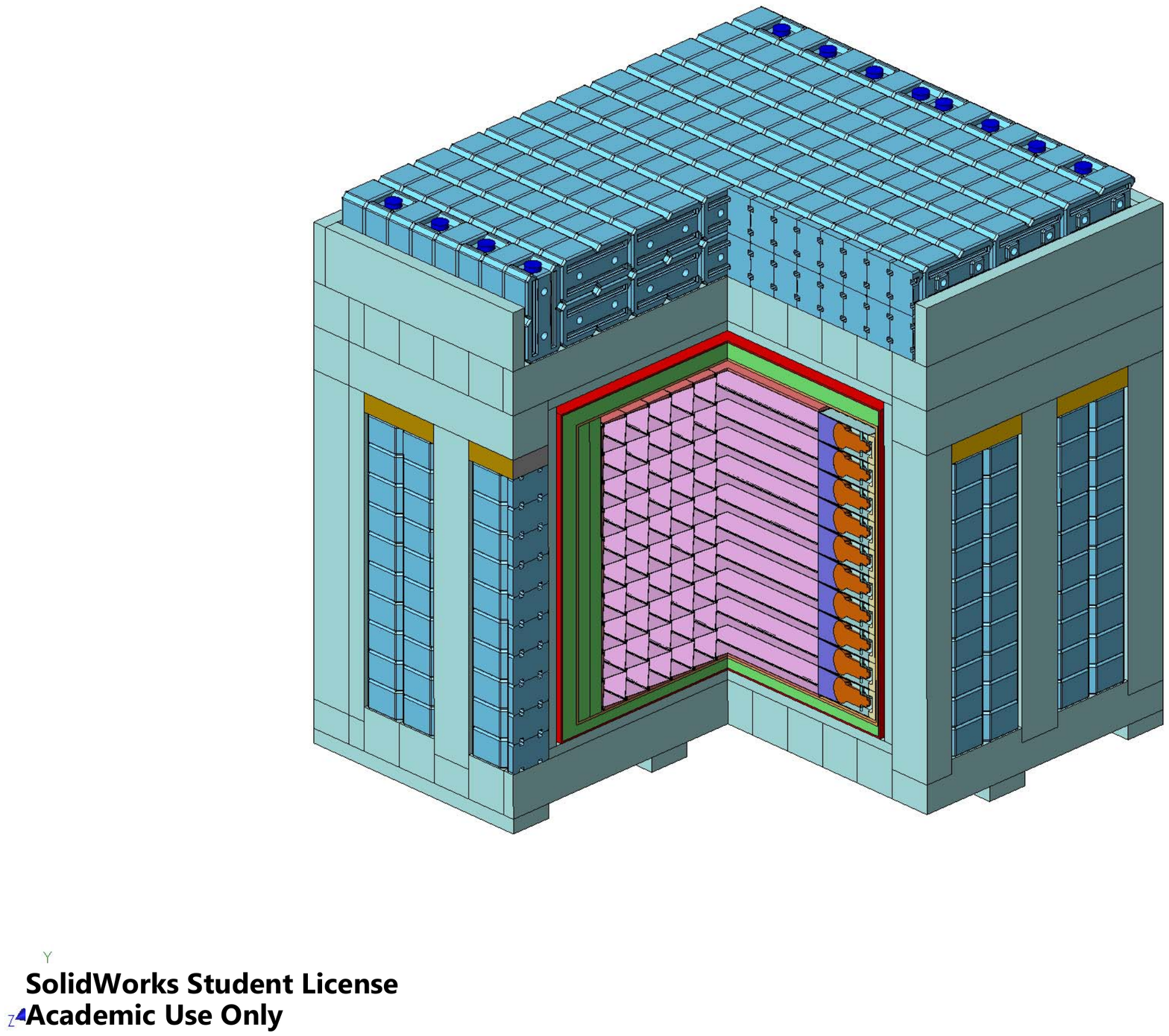}
\caption{Phase I detector.}
\end{subfigure}
\begin{subfigure}[b]{0.523\textwidth}
\includegraphics[clip=true, trim=0in 1.15in 0in 0in,  width=\textwidth]{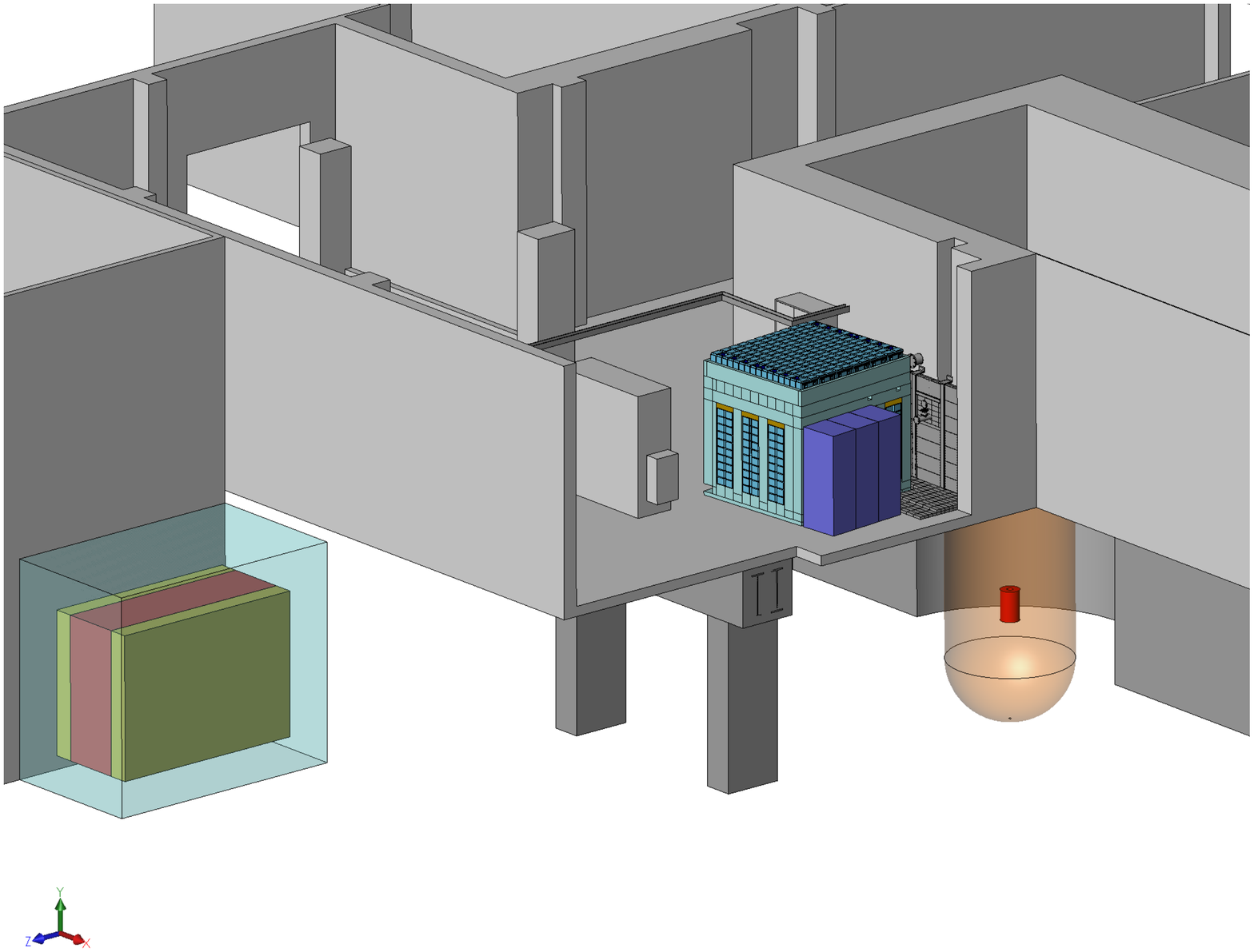}
\caption{Detector placement at HFIR.}
\end{subfigure}
\caption{(a) The inner structure of the Phase I detector.  
The segmented inner volume can be seen with a PMT readout.  
The shielding package consists of lead, polyethylene, and water.
(b) The positions of the Phase I (near) and Phase II (far) detectors at the HFIR complex relative to the reactor core (red cylinder).}
\label{fig:Detector}
\end{figure}

The PROSPECT experiment has been planned in phases \cite{Ashenfelter:2013}.
This has several benefits as it will be able to address experimental situations in a timely manner, to mitigate any risks that arise, to provide systematic controls and increased physics reach, and allow flexibility in response to results from Phase I.
The Phase I detector will be located $\approx 7$ m from the reactor core, as seen in Figure \ref{fig:Detector}.
It consists of a 12 x 10 grid of optically separated segments that are 15 cm x 15 cm x 1.19 m.
These segments are filled with ${}^6$Li-loaded liquid scintillator.
The optical separators are reflective in order to efficiently collect light to attain a $4.5\%/\sqrt{\textrm{E}}$ energy efficiency at 1 MeV.
At each end of the segment are 5 inch photomultiplier tubes (PMTs).
In order to extend the baseline coverage of the Phase I detector, a Phase I+ is planned where the entire Phase I detector in its shielding package is moved away from the reactor core giving a total coverage from 7-12 meters.
Logistical and engineering considerations for multiple Phase I detector locations have been confirmed with HFIR engineers.
This will provide greater physics reach through an increased L/E range and the possibility for systematic cross-checks by comparing detector data at the same baseline measured in different detector segments.
In addition to this movable detector, a later addition of a Phase II detector will be situated $\approx$ 18 m from the reactor core, which will provide greatly enhanced coverage over the low $\Delta m_{14}^2$ parameter space.
Since PROSPECT will be conducted in phases, the collaboration can respond to the information gained in Phase I and adjust the design of Phase II in order to perform a more precise measurement.

\section{PROSPECT Physics Reach}

\begin{figure}[htb]
\centering
\includegraphics[clip=true, trim=0in 0in 0in 0in, width=0.7\textwidth]{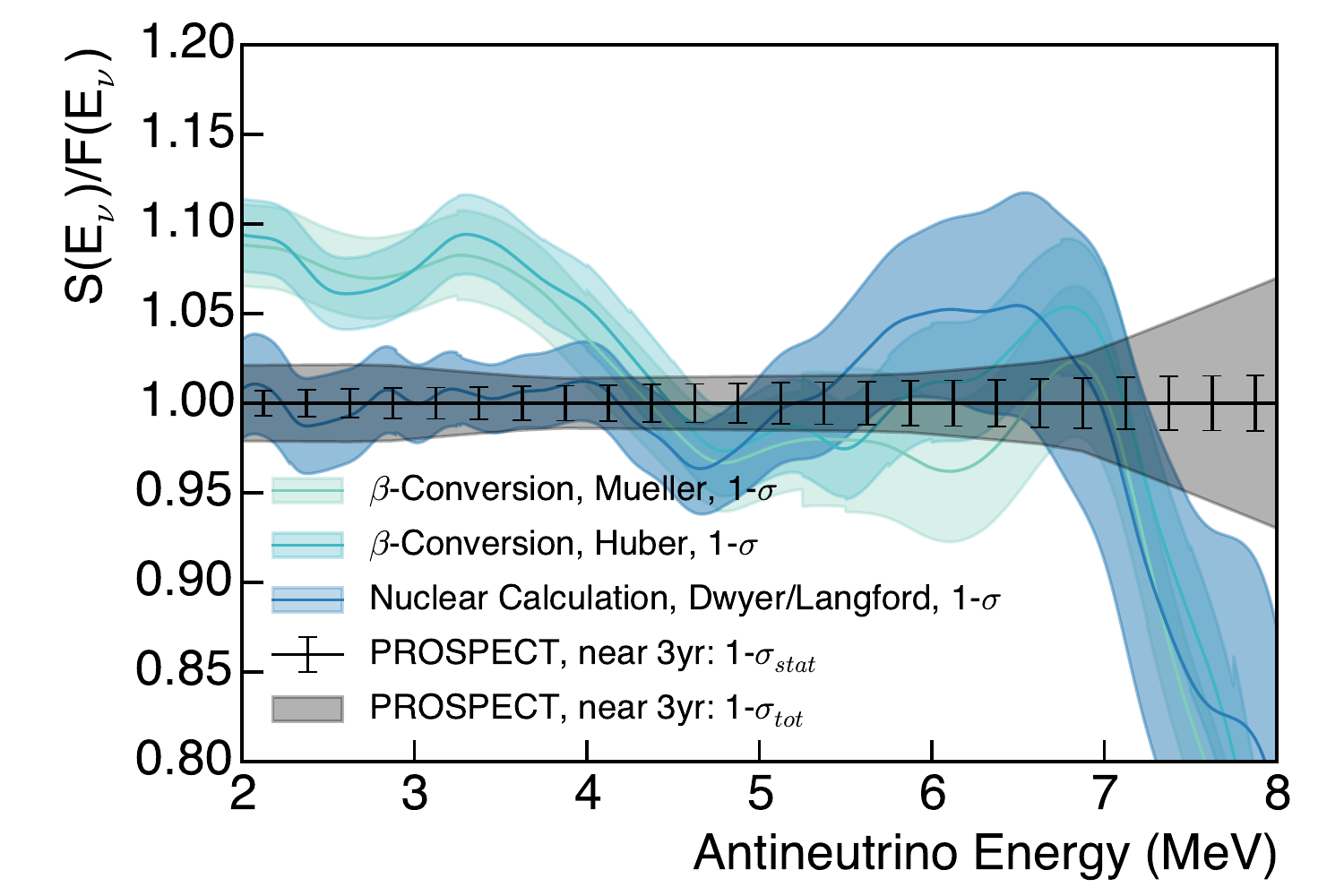}
\caption{Three models of the ${}^{235}\textrm{U}$ $\bar{\nu}_e$ energy spectrum are shown relative to a smooth approximation of a spectrum to be measured by PROSPECT.}
\label{fig:Models}
\end{figure}

PROSPECT will measure the reactor antineutrino energy spectrum with a precision sufficient to be able to distinguish between several common models (Figure \ref{fig:Models}).
This precise measurement  of the single component spectrum of a ${}^{235}\textrm{U}$ fueled reactor will directly constrain models which, in turn, will aid experiments probing the neutrino mass hierarchy and improve existing and future measurements by reducing uncertainties on the reactor $\bar{\nu}_e$ background.
In addition, there are further applications in the field of nuclear nonproliferation and in the applied nuclear physics community.


\begin{figure}[htb]
\centering
\includegraphics[ width=0.7\textwidth]{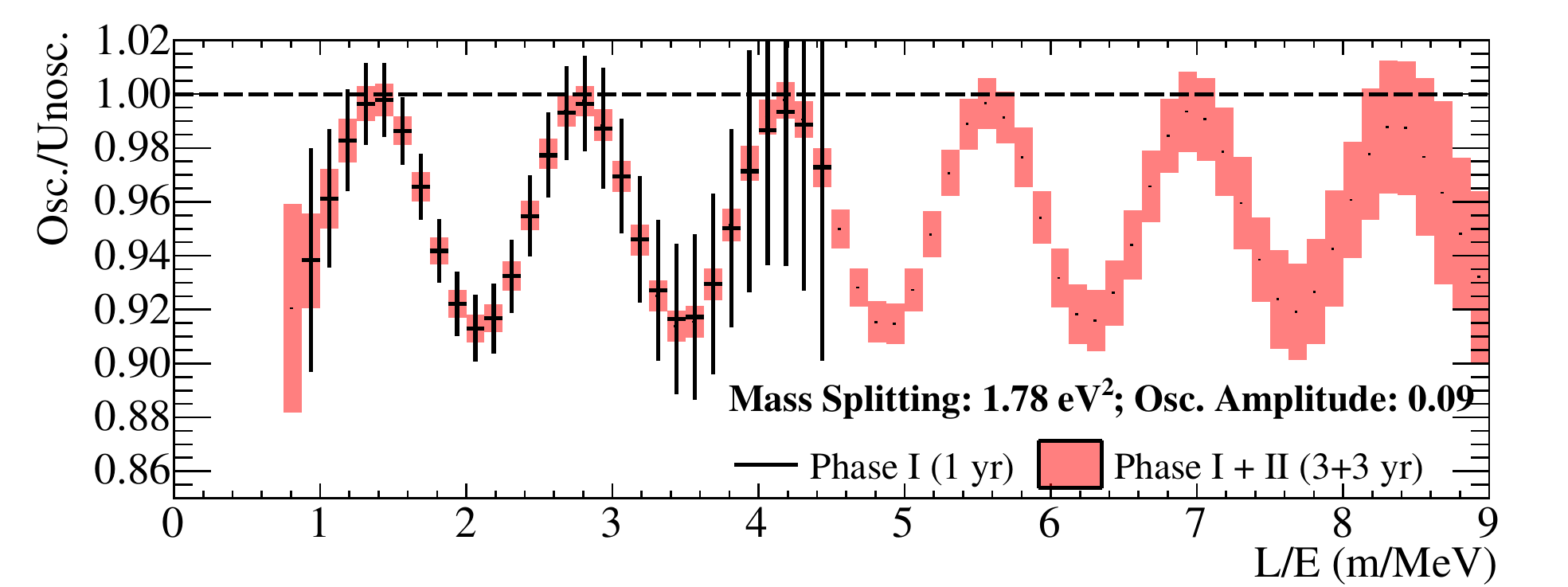}
\caption{(The ratio of the L/E distributions of the oscillated spectrum divided by the unoscillated spectrum using the Kopp best fit point.
The depth of the oscillation shown in the figure is determined by $\sin^2 2\theta_{14}$ and the frequency is determined by $\Delta m^2_{14}$.}
\label{fig:LoverE}
\end{figure}

\begin{figure}[htb]
\centering
\includegraphics[width=0.7\textwidth]{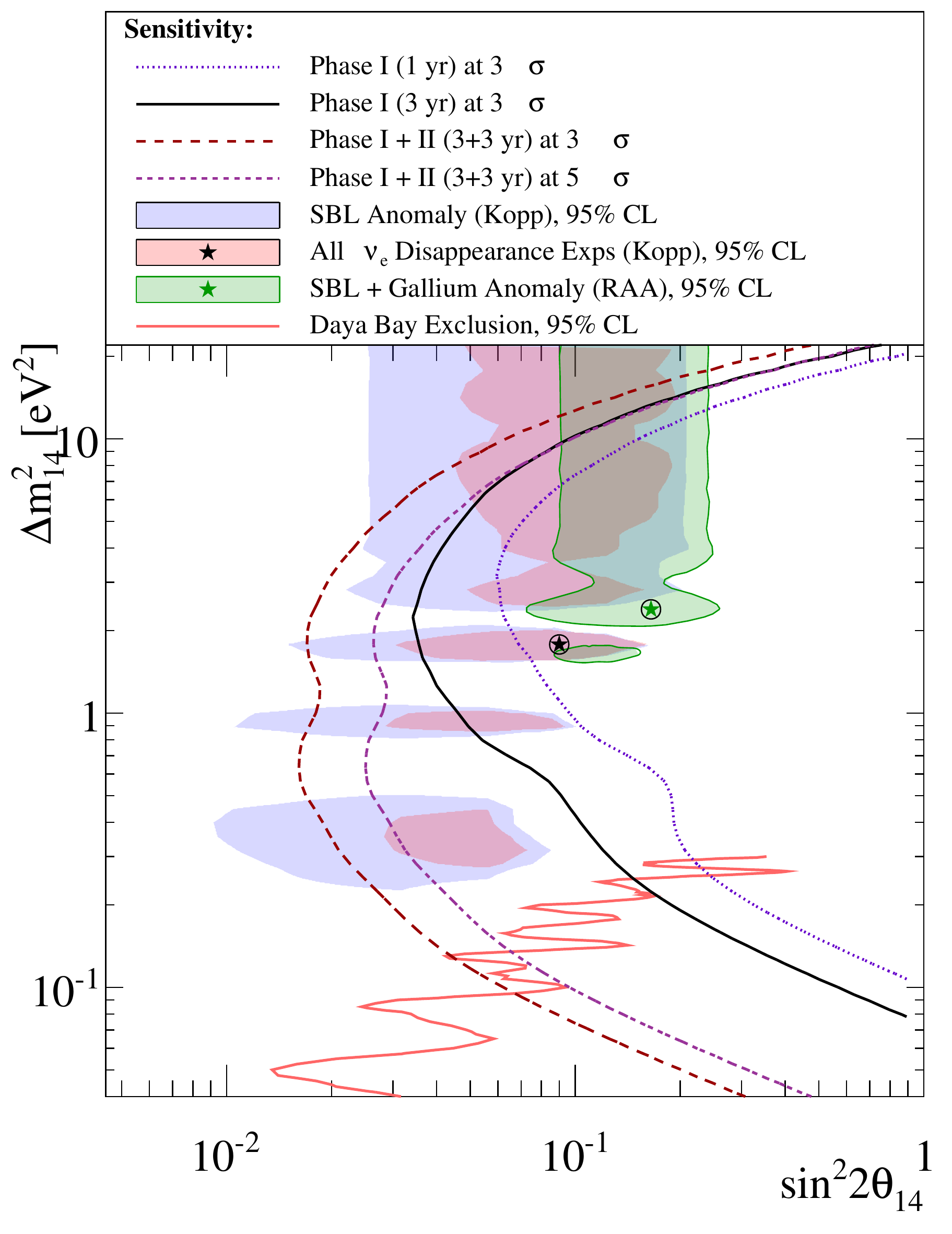}
\caption{The predicted sensitivity reach into the $\Delta m^2_{14}-\sin^2 2\theta_{14}$ parameter space.
Shown are the 3$\sigma$ sensitivity curves for Phase I at 1 and 3 years of data taking and the 3$\sigma$ and $5\sigma$ curves for Phase II at 3 years of data taking in addition to 3 years of Phase I data taking.}
\label{fig:Osc}
\end{figure}

The sterile neutrino search will be a relative measurement between the segments of the detector, and therefore will be totally independent of an absolute spectrum model prediction.
All inputs for the sensitivity calculations are based either on direct prototype measurements at HFIR or on data-benchmarked PROSPECT simulations.
Input parameters of primary importance are: 4.5\%/$\sqrt{\textrm{E}}$ energy resolution, 15 cm position resolution, 1:1 signal to background ratio at the front position of the Phase I detector, and 41\% reactor up-time.

The sensitivity of PROSPECT is shown in Figure \ref{fig:Osc} for the different planned phases of the experiment.
An example of the L/E distribution for the Kopp best fit point is shown in \ref{fig:LoverE}.
The blue and red shaded regions represent the best fit sterile oscillation parameter space, as calculated by Kopp \cite{Kopp:2013}, of the short baseline reactor anomaly and the global fit from all $\nu_e$ disappearance experiments, respectively.
The green shaded region is the anomalous region according to the Light Sterile Neutrino whitepaper (RAA) fit \cite{RAA:2012}.
Currently, there is little consensus on the appropriate global fit to compare with.
Although these fits have many similarities, there are a several differences that lead to significant changes in the allowed parameter space.
Each fit includes constraints from the Bugey3 spectrum shape, but RAA also includes the ILL spectrum shape which shifts the anomalous region to the higher $\Delta m_{14}^2$ and higher $\sin^2 2\theta_{14}$.
Both fits use information from the SAGE and GALLEX experiments, but with slightly different cross sections and thus different corrected rates.
Lastly, when calculating the best-fit region for all $\nu_e$ disappearance experiments, Kopp included limits from long baseline reactor experiments, carbon-12 experiments, and solar experiments.
All of these differences contribute to a visible divergence between the 95\% confidence level curves of the global fits.
PROSPECT has been designed to cover the Kopp best-fit point with a single year of data in Phase I at $3\sigma$ and cover the majority of the RAA parameter space at $5\sigma$ in the same period.
It will also cover the majority of all suggested parameter space at $5\sigma$ though the combination of Phase I and Phase II.



\section{Conclusion}
PROSPECT will make a precise measurement of the ${}^{235}\textrm{U}$ spectrum which will provide new constraints on reactor antineutrino emission models.
This measurement will prove complimentary to current and future measurements at LEU fueled reactors.
PROSPECT will also perform a search for a sterile neutrino on the $1 \textrm{eV}^2$ scale and, within one year of Phase I data taking, have 3$\sigma$ coverage over the Kopp global best fit.
These studies of $\bar{\nu}_e$ disappearance will be complementary to the current Fermilab short baseline program which will focus on a search for sterile neutrinos through the measurements of $\nu_\mu$ to $\nu_e$ appearance and $\nu_\mu$ disappearance.

\Acknowledgments
This material is based upon work supported by the U.S. Department of Energy Office of Science.
Additional support for this work is provided by Yale University and the Illinois Institute of Technology.
We gratefully acknowledge the support and hospitality of the High Flux Isotope Reactor and the Physics Division at Oak Ridge National Laboratory, managed by UT-Battelle for the U.S. Department of Energy.

\end{document}